\documentclass{moriond}

\def\Journal#1#2#3#4{{#1} {\bf #2}, #3 (#4)}

\def\be{\begin{equation}}
\def\ee{\end{equation}}
\def\bea{\begin{eqnarray}}
\def\eea{\end{eqnarray}}

\newcommand{\Photo}{\includegraphics[height=35mm]{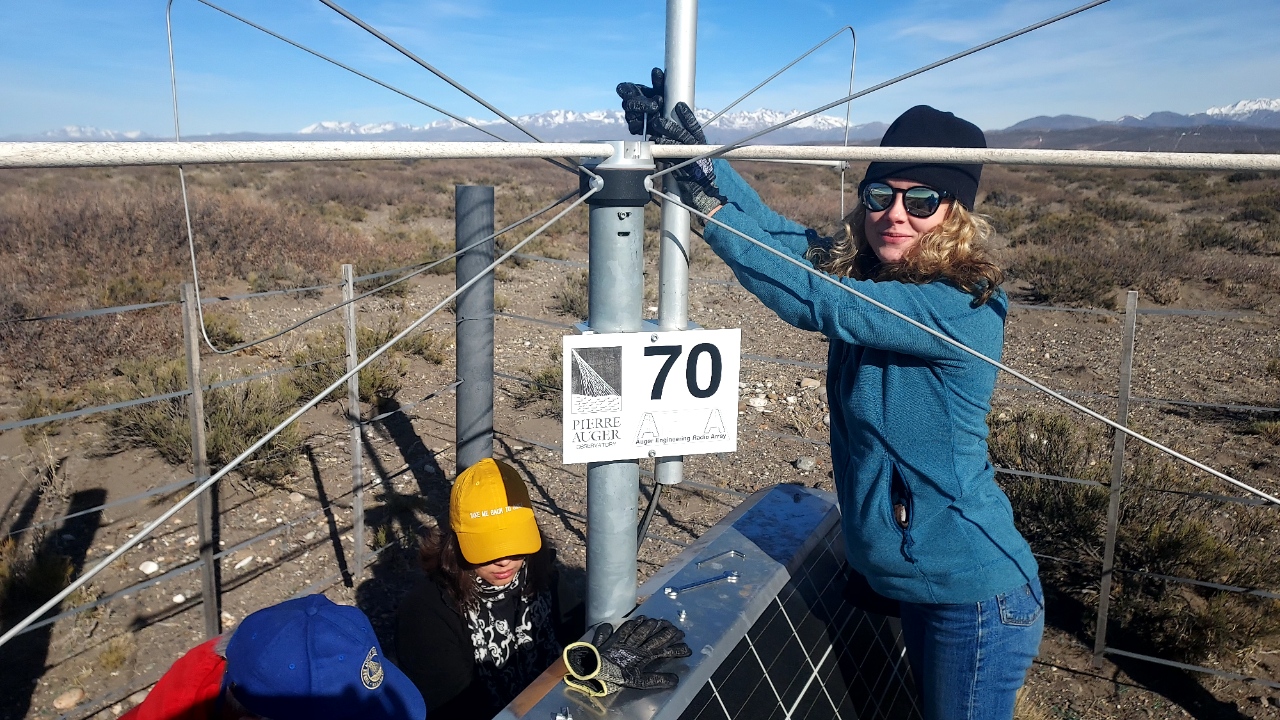}}

\begin{document}
\vspace*{4cm}
\title{The Giant Radio Array for Neutrino Detection (GRAND) and its prototype phases}

\author{Marion Guelfand, on behalf of the GRAND collaboration}

\address{Sorbonne Université, CNRS, Laboratoire de Physique Nucléaire et des Hautes Energies (LPNHE),\\
4 Pl. Jussieu, 75005 Paris, France\\
Sorbonne Université, CNRS, Institut d'Astrophysique de Paris,\\
98 bis bd Arago, 75014, France}

\maketitle\abstracts{Multi-messenger astronomy is key today to broaden our understanding of the high energy Universe. When an ultra-high energy (UHE) particle interacts in the atmosphere or underground, it initiates an extensive air shower that produces a coherent radio emission in the 50-200 MHz  range. The GRAND project is an envisioned observatory of UHE particles (cosmic rays, gamma rays and especially neutrinos) that will consist of 200,000 radio antennas deployed over 20 different locations each of $\sim$ 10,000 $\rm km^2$.
In its current phase, it consists of 3 prototypes running autonomously in 3 different locations: GRAND@Auger in Argentina, GRAND@Nançay in France and GRANDProto300 in China, all at commissioning stage. The first goal of these pathfinders is to demonstrate the viability of the GRAND detection concept. GRANDProto300 will also propose a rich science case, by allowing the study of the transition between galactic and extra-galactic cosmic ray sources. In the following, we present the detection concept, the preliminary designs and layout and the ongoing developments of the experiment.}

\section{Introduction}

The Giant Radio Array for Neutrino Detection (GRAND) is an envisioned observatory which aims to detect the radio signals induced in the Earth's atmosphere by ultra-high energy (UHE) particles (cosmic rays, gamma rays and neutrinos). It will consist of 200,000 radio antennas over 200,000 $\rm km^2$, divided into 20 sub-arrays worldwide. In the current status, three prototype detectors, which serve as pathfinders, are operating autonomously: GRANDProto300 in China, GRAND@Auger in Argentina and GRAND@Nançay in France. Below, we will present the detection principle of the experiment, the setup and the goals of the prototyping arrays, their ongoing developments, and the next steps that are planned to be achieved.

\section{Detection concept}

When a UHE particle enters the atmosphere, it interacts with an air molecule and produces a cascade of secondary particles, called an extensive air shower (EAS), that generates an electromagnetic emission mainly due to the geomagnetic effect~\cite{Scholten_2007}. UHE tau neutrinos can also generate such electromagnetic signals by interacting in the Earth's crust, which is a denser medium. A tau lepton is produced, which exits the rock and decays in the atmosphere, generating an EAS~\cite{Fargion_2002}. The radio emission produced by UHE particles is coherent in the [10-200] MHz frequency range and leaves a radio footprint on the ground, with an amplitude high enough to detect EAS with energy $ > 10^{16.5}\, \rm eV$~\cite{Huege_2016}.

The design of GRAND will enable the detection of cosmic ray air showers with large zenithal inclinations ($> 70^{\circ}$) and Earth-grazing neutrinos~\cite{GRAND_2020}, whose radio footprint can reach tens of km due to projection effects of the signal on the ground and the increasing distance between the emission region and the antennas. By targeting these EAS, GRAND will enable a good sampling of the radio signal with a sparse density of radio antennas, allowing the deployment of detectors over gigantic areas and increasing its sensitivity. 

\section{Prototypes}

Three prototype detectors, still in commissioning, are currently operating simultaneously. Their main goal is to validate the GRAND detection principle i.e., demonstrate that for very inclined EAS, radio antennas can be triggered autonomously (with radio signal alone and without using any external detection system) with high efficiency and excellent background rejection.

\subsection{Detector setup}
The detector's central component, called HorizonAntenna, is designed to have an optimal sensitivity for near-horizontal signals~\cite{GRAND_2020}. It is a bow-tie antenna with 3 perpendicular arms, allowing the detection of the 3 polarizations of the electromagnetic radiation, positioned at a height of $3.5\,\rm m$ above the ground and optimized for the 50-200 MHz frequency range. The upper limit of 200 MHz is chosen to be able to detect the Cherenkov cone and because the signal-to-noise ratio (SNR) is higher at larger frequencies. Each antenna is self-powered thanks to a $150\, \rm W$ solar panel and the signal from each polarization direction is amplified by about 23 dB by a Low Noise Amplifier, before being transported to the front end board (FEB), directly placed in a box at the bottom of the antenna. An electronic treatment of the signal is then performed. The analog signal is first filtered in the 50-200 MHz frequency band and it is then digitized using a 14-bit ADC running at a sampling rate of 500 Megasamples/s. Digitized data are then processed inside a FPGA and read by the CPU. At this stage, trigger is also performed~\cite{Martineau_Moriond2024} and data are then transmitted to the Central Data Acquisition (DAQ) by WiFi, which is designed to be able to manage a $10\, \rm Hz$ rate per detection unit for trace transfer.

\subsection{Goals and layout}
The locations of the sub-arrays, including prototypes, are chosen in radio-quiet environments, which is an essential feature for autonomous detection of EAS.

Ten antennas were deployed at the Pierre Auger Observatory site, in Malargue, Argentina, between March and November 2023, using Auger infrastructure. GRAND@Auger is an ideal testbench to evaluate the quality of the reconstruction procedure, in terms of arrival direction, energy and nature of the primary particle. Indeed, with an expected rate of around 1 cosmic ray EAS/day in coincidence with Auger SD data, it will allow calibration and validation of the reconstruction event by event.

\begin{figure}[!h]
         \includegraphics[width=0.44\columnwidth]{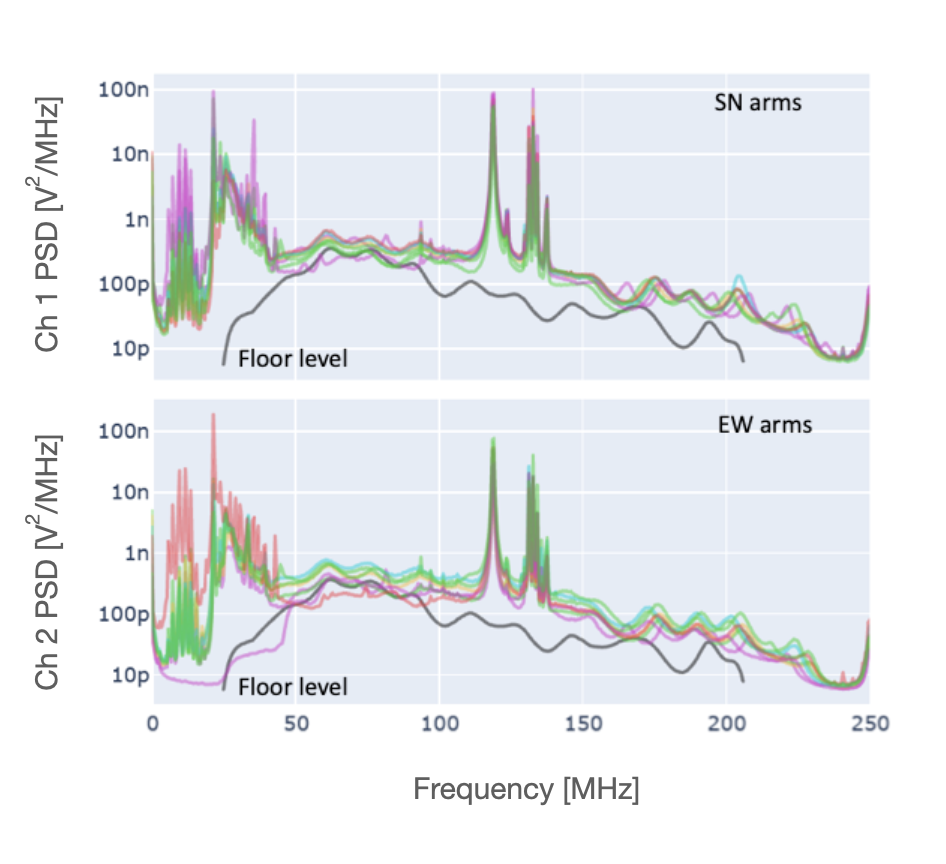}
         \includegraphics[width=0.5\columnwidth]{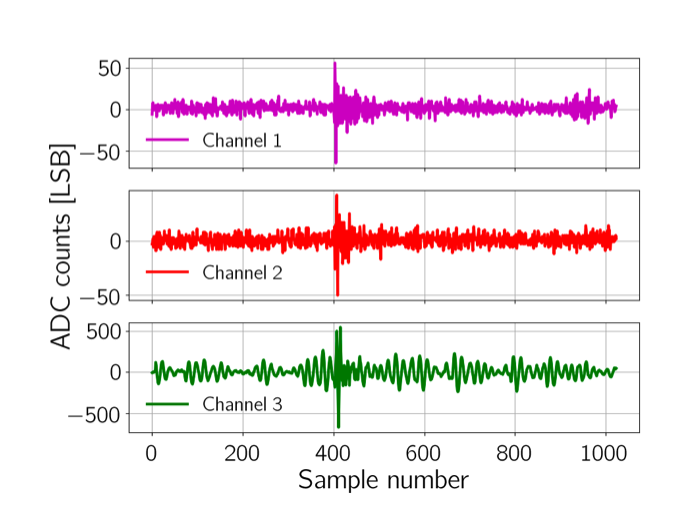}
    \caption{\textit{(Left)} Power spectrum (PSD) measured at GRANDProto300 on different units and for two polarization directions: South-North and East-West. The frequency spectrum is not filtered for all detection units. It is nominal with known background emitters at given frequencies (short radiowaves, TV carrier). The floor level corresponds to the mean simulated galactic noise computed at DAQ input. \textit{(Right)} Transient event measured at GRANDProto300 on one antenna: channel 1=North/South, channel 2=East/West, channel 3=vertical.}
      \label{fig:transient}
\end{figure}

Additionally, thirteen detection units -which constitute the first phase of the GP300 detector- were deployed at Xiao Dushan, Dunhuang, China, in February 2023~\cite{Ma_ICRC2023}, where the topography is particularly favorable for neutrino detection due to the presence of mountains. For an optimal topography, two opposite mountain ranges separated by a few tens of kilometers would be needed. One range would serve as a target for neutrino interaction, while the other would act as a screen on which the created radio signal is projected. From this initial setup, data are being collected and analyzed in order to validate the detector unit design, which serves as a foundation for the complete GRANDProto300 array (see Figure~\ref{fig:transient}). The reconstruction procedure of the arrival direction is also under validation, using the signal emitted by the beacon antenna placed on the roof of the central station.

FInally, four detection units were deployed at the Nançay Radio Observatory in France during autumn 2022. The primary objective of GRAND@Nançay is to serve as a testbench for hardware and triggering~\cite{Correa_ICRC2023}.

For the three prototypes, still at commisionning phases, the noise level is within specifications and the first transient signals are detected.  

\section{Future GRAND timeline}
\subsection{GRANDProto300: next steps}
\begin{figure}[!h]
         \includegraphics[width=0.5\columnwidth]{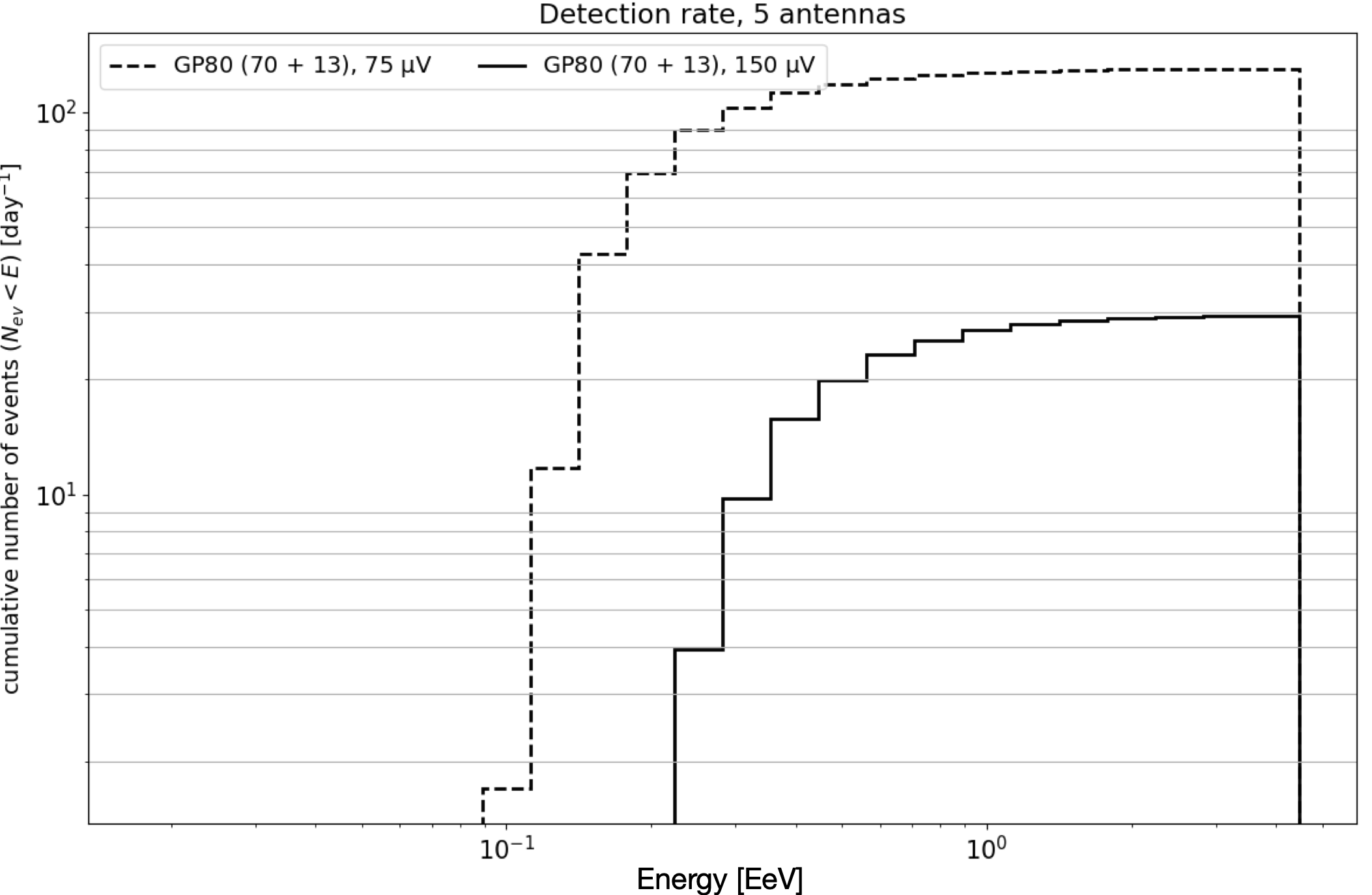}
        \includegraphics[width=0.5\columnwidth]{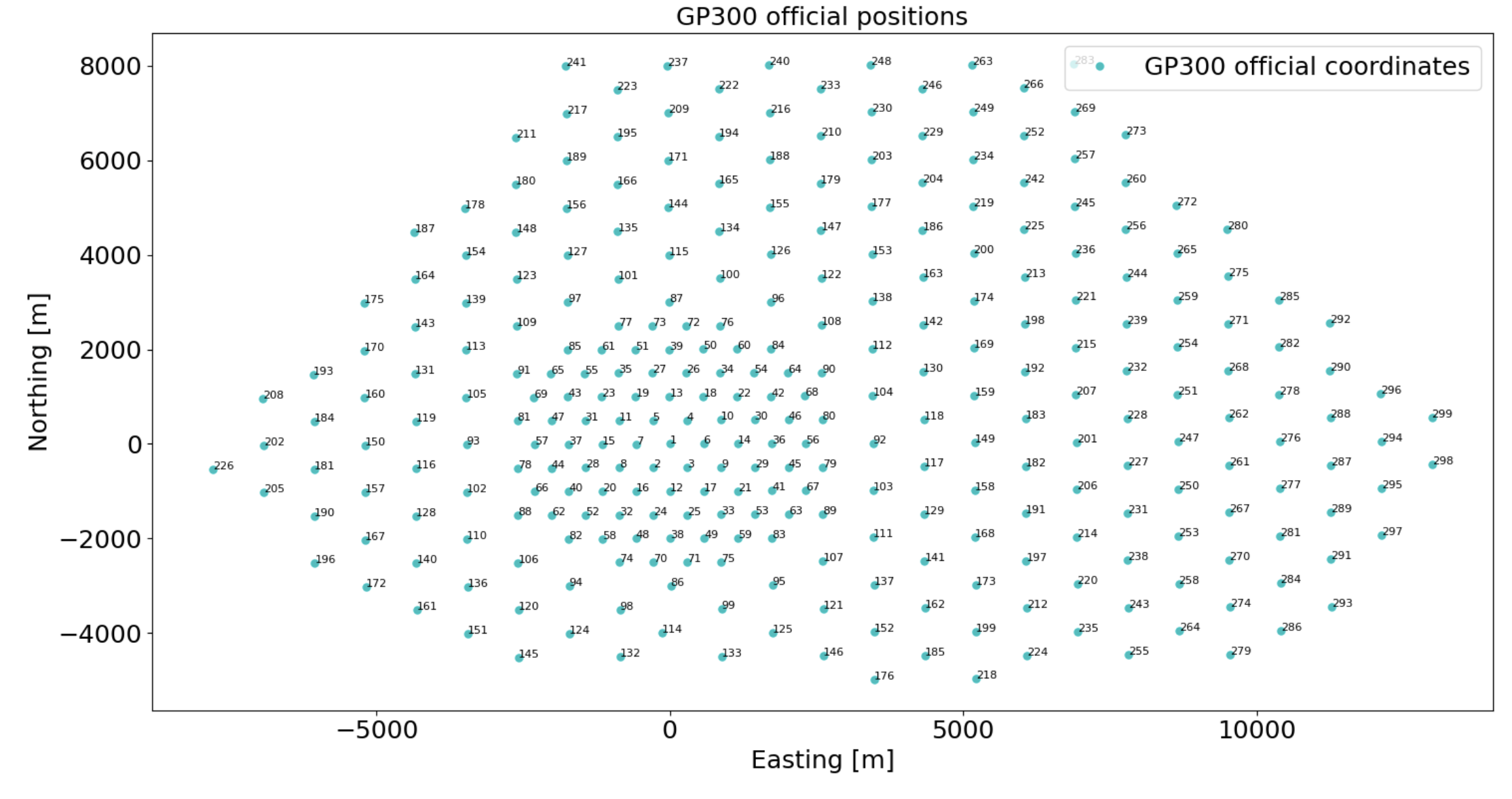}
    \caption{\textit{(Left)} GP300 differential event rate as a function of the primary cosmic ray energy for one preliminary layout with 13+70 antennas each separated by a distance of 1000m. The trigger conditions require a minimal number of 5 antennas with a voltage signal peak-to-peak above $5\sigma = 75\rm  \mu V/m$ (dotted line) or above  $10\sigma = 150\rm  \mu V/m$ (full line). \textit{(Right)} Complete GP300 layout with infill.}
      \label{fig:layout}
\end{figure}

Following administrative approval obtained in May 2024, 70 more detection units already built will be deployed on the GRANDProto300 site in Autumn 2024, and will allow to detect cosmic rays. Various designs are still under consideration to obtain a good balance between detected event rates, reconstruction quality and to prepare the design of the antenna arrays in future phases of the project. An expected differential rate derived from ZHaireS simulations is of 30 cosmic ray EAS/day with conservative trigger settings (ten times the minimum noise level) and a layout with an antenna separation of 1,000$\,\rm m$ (see Figure~\ref{fig:layout} left). The cosmic ray data will be used to measure the detection efficiency and the background rejection efficiency, and hence validate the GRAND detection principle. 

The last step of this prototype phase will be the construction and the deployment of the remaining 200 units to form the complete GRANDProto300 detector (see Figure~\ref{fig:layout} right). This array will have the capability to investigate cosmic rays within the energy range of $10^{16.5}$ to $10^{18}$ eV, where the transition between galactic and extra-galactic  cosmic ray sources is expected~\cite{Dawson_2017}. The high statistics events of GRANDProto300 will allow accurate measurements of the energy, composition and distribution of the arrival direction of cosmic rays to distinguish between different astrophysical source models. The array will also allow for the detection of radio transients, such as Fast Radio Bursts.

\subsection{The road to neutrino astronomy}

It is expected that the design of the detector units will be finalized by 2028, to conduct the second stage of the GRAND project: GRAND10K North/South, that will lead to the construction of two arrays of 10,000 antennas each. The two candidate sites are based in China and in Argentina and will serve to test challenges related to large-scale arrays, such as data transfer, storage, communication. With its sensitivity and its full sky coverage, GRAND10K North/South will have the discovery potential for EeV neutrino. Once operational, the design of the detection units will be frozen and duplicated following an industrial approach.

\end{document}